\documentclass[aps,showpacs,preprintnumbers,twocolumn]{revtex4-2}

\usepackage[centertags]{amsmath}
\usepackage{amsfonts}
\usepackage{amssymb}
\usepackage{amsthm}
\usepackage{newlfont}
\usepackage{graphicx}
\input{epsfx}

\begin{document}

\title{Consistency of quantum computation and the equivalence principle}
\author{Marcin Nowakowski\footnote{Electronic address: marcin.nowakowski@pg.edu.pl}}
\affiliation{Faculty of Applied Physics and Mathematics,
~Gdansk University of Technology, 80-952 Gdansk, Poland}
\affiliation{National Quantum Information Center of Gdansk, Andersa 27, 81-824 Sopot, Poland}


\begin{abstract}
The equivalence principle, being one of the building blocks of general relativity, seems to be also crucial for analysis of
quantum effects in gravity.
In this paper we consider the question if the equivalence principle has to hold for consistency of performing quantum computation in gravitational field. We propose an analysis with a looped evolution consisting of steps both in the gravitational field and in the accelerated reference frame. We show that without the equivalence principle the looped quantum evolution cannot be unitary and looses its consistency. For this reasoning the equivalence principle is formulated in terms of the gauge transformations and is analyzed for particles acquiring an appropriate phases associated with the actions over the looped path. In consequence, to keep consistency of quantum operations in gravitational field, it is required to keep some quantum variant of the equivalence principle. This proves importance of the quantized versions of this fundamental gravitational principle for quantum information processing.

\end{abstract}

\maketitle

The equivalence principle is one of the foundations of general relativity. Although its roots are classical, the long-standing research is focused on inclusion of this principle \cite{Aharonov1, Aharonov2, Marletto, Bose, Marletto2, Marletto3, Hardy, Zych1} into the future  quantized version of gravity which is based on the assumption that it should be also a foundation of quantum gravity. There has been also a long-standing debate if gravity itself is quantum but recent research \cite{Bose, Bose2} shed some light on the quantumness of gravity proposing interferometric experiments testing its quantum nature. In this context,  it is an objective of this paper to consider the issue of consistency of distributed quantum information processing in background of weak gravitational fields which can be considered quantum for this task \cite{Feynman}.

\begin{figure}[h]
\centerline{\includegraphics[width=9.5cm]{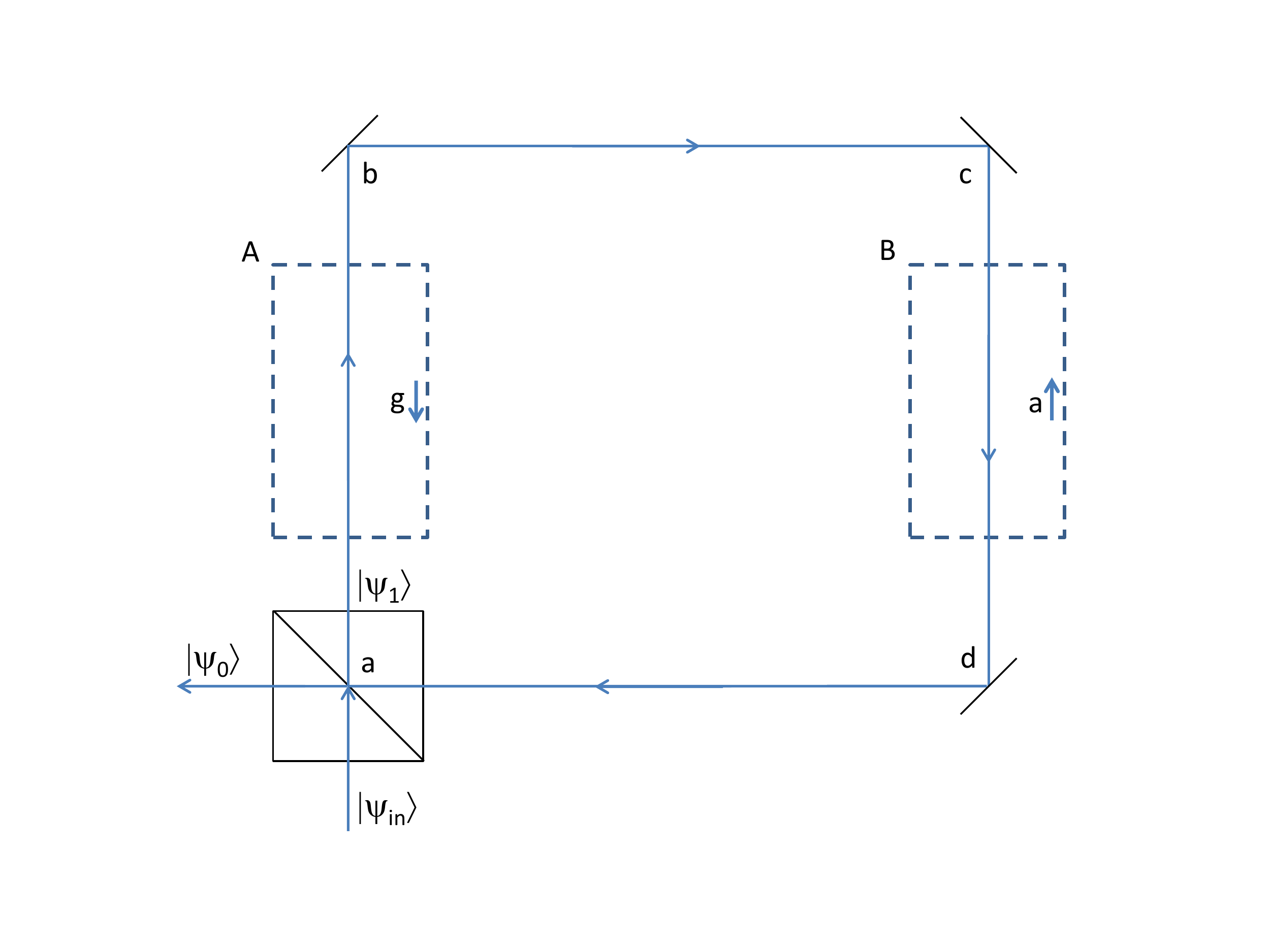}}
 \caption[Mach-Zehnder Interferometer]{A looped quantum-gravitational interferometer with space-time region A in gravitational field g and space-time region B with acceleration \textbf{a}=-\textbf{g}. The looped particle traverses the loop $abcda$.}
 \label{Mach-Zender Interferometer}
\end{figure}

Our proposal is based on the model of a looped system (Fig. \ref{Mach-Zender Interferometer}) with quantum evolution connecting two different regions of space-time from which one is in gravitational field and the another one being a gravity-free region is only accelerated. For this consideration we engage a standard understanding of the equivalence principle assuming that \textit{the state of motion of a point particle at rest in a local uniform gravitational field \textbf{g} is indistinguishable from the state of motion of a point particle in an accelerated reference frame (with $\textbf{a}=\textbf{-g}$) in a gravity-free region}. Alternatively, it can be formulated in terms of local reference frames stating that there is no way, by experiments confined to infinitesimally small regions of space-time, to distinguish one local Lorentz frame in one region of space-time from any other local Lorentz frame in the same or any other region \cite{Gravitation}.  This formulation is a basis for recent \cite{Hardy} quantum proposal of the equivalence principle with quantum coordinate systems and is naturally connected with the problem of defining a quantum reference frame \cite{Brukner, Aharonov3} and relativity of quantum superpositions \cite{Zych2}.

In the context of this paper, it is fundamental to note that gauge transformations govern the way of transforming a system from a region with gravitational field $\textbf{g}$ to the accelerated reference frame with $\textbf{a}=\textbf{-g}$.
In quantum mechanics the Schrodinger equation is invariant when both potentials and the wave-function are subject of these transformations. For the vector potential gauge transformations are:
\begin{equation*}
A'_{\mu}=A_{\mu}+\partial_{\mu}X
\end{equation*}
Then the quantum state is transformed as: $|\Psi'\rangle=U|\Psi\rangle=e^{i\phi}|\Psi\rangle$ with the gauge parameter $U$ being a phase factor which is actually an element of the unitary group $U(1)$ for Maxwell theory. It is also worth mentioning that this element is space-time dependant which is vital for further reasoning. For a series of gauge transformations $U_{1}\rightarrow U_{2}\ldots \rightarrow U_{n}$ of the state $|\Psi\rangle$ the composition law holds:
\begin{equation}
|\Psi\rangle \rightarrow U_{1}|\Psi\rangle \rightarrow U_{2}U_{1}|\Psi\rangle\rightarrow \ldots \rightarrow U_{n}\ldots U_{2}U_{1}|\Psi\rangle
\end{equation}
For our example of the closed quantum-gravitational loop we get effectively a version of the Wilson loop (being also gauge invariant)  for the closed space-time path. The gauge parameter for this case is periodic ($U(\lambda + 2\pi)=U$ for some $\lambda$ parametrizing the path) on the loop which physically is in line with the equivalence principle and with the conservation of energy as presented in this paper.

For the linearized gravitational  fields \cite{Gravitation} the Riemann tensor is gauge invariant with the same property inherited by the stress-energy tensor and Einstein tensor. Thus, we can formally further consider quantum-gravitational systems having this property. However, it is an open question if the following reasoning is also fully applicable to gravity with non-linear gravitational field equations.


We shall analyze now the model of the looped evolution and show that if the quantum version of the equivalence  principle, associating quantum phase with each path in different background fields, does not hold, then the looped evolution cannot be unitary. We consider further a closed loop (Fig. 1) with part of the loop (space-time region A) being in the gravitational field $g$ and part in the accelerated system with $a=-g$ (space-time region B).
We follow the Feynman path methods assuming that each sub-path contributes to the final probability amplitude of a quantum process adding actions for paths in regions A and B. By the very property of associating a quantum phase with a path traversed by a quantum system in gravitational field, we assume quantumness of the interaction which is actually also a hidden assumption in this type of analysis and we also assume consistency of the single path traversing both regions.

The probability amplitude for the history branch of the 1-loop process is:
\begin{equation}
    \mathcal{A}_1=\langle \Psi_{1}|e^{\frac{i}{\hslash}S_{da}}e^{\frac{i}{\hslash}S_{cd}}e^{\frac{i}{\hslash}S_{bc}}e^{\frac{i}{\hslash}S_{ab}} |\Psi_{1}\rangle
\end{equation}
For simplification we can consider only two paths of this loop: $a\rightarrow c$ (including $b$) and $c\rightarrow a$ (including $d$) putting the actions $S_{ac}$ and $S_{cd}$. Thus, a quantum particle traversing this loop acquires phase $e^{i\phi_{ac}}$ associated with existence of the gravitational field $g$ on the sub-path $a\rightarrow b$ and $e^{i\phi_{ca}}$ with traversing the accelerated (by $a$) sub-path $c\rightarrow d$. Following one gets:

\begin{equation}
    \mathcal{A}_1=\langle \Psi_{1}|e^{i\phi_{ca}}e^{i\phi_{ac}} |\Psi_{1}\rangle
\end{equation}
where $\phi_{ca}=S_{ca}/\hslash$ and $\phi_{ac}=S_{ac}/\hslash$ (we omit in the calculation the acquired phases of paths $b\rightarrow c$ and $d\rightarrow a$ which is dependant on a particular implementation of the system).

Further for the n-loop process its probability amplitude is:
\begin{equation}
    \mathcal{A}_n=\langle \Psi_{1}|[e^{i\phi_{ca}}e^{i\phi_{ac}} ]^n|\Psi_{1}\rangle
\end{equation}
which is in agreement with the composition law for the gauge transformations $U=[U_2U_1]^n=[e^{i\phi_{ca}}e^{i\phi_{ac}} ]^n$.

Even not assuming correctness of the gauge transformation on the closed loop (closed history associated with the evolution of the particle), one would be forced to introduce a relation between systems A and B (the relation actually is created due to existence of the connection created by the history of a particle traversing both 'regions' A and B) which is related to changes in the state of the system. As an implication of this reasoning the change of state (associated with the acquired phase) within A has to be related to the change on B, which is imposed  by the consistency of the traversing particle's history. The binding evolution of the traversing particle is unitary which is postulated by quantum mechanics and is a necessary condition of consistency of quantum  information processing. If that does not hold, then many paradoxes would arise like those related to  contradiction of the quantum no-cloning principle.

There are two important principles necessary for ensuring unitarity of a quantum evolution with gravitational field interaction for our model:

1. Firstly, we assume that laws of physics apply in the same way to each history branch which has its formulation in 1-loop, 2-loop up to n-loop processes where we apply the superposition rule up to the n-loop processes imposing its application to gravitational fields.

2. Secondly, we assume correctness of the equivalance principle imposing a direct relation between the gravitational field and the accelarated system. This relation is being ensured by the binding history of the quantum particle traversing the system A and then system B which also creates information connection between the regions of space-time with different gravitational characteristics \cite{Nowakowski}.

It is now crucial to show that the processes are unitary up to the n-loop processes in the infinite regime with convergence of the superposition series. For the $0$-loop process where the particle is just reflected from the BS (beam-splitter), one gets:
\begin{equation}
    |\Psi_{in}\rangle \rightarrow |\Psi_{out}\rangle= |\Psi_{in}\rangle
\end{equation}
For the 1-loop process, where a particle traverses the loop only once but acquiring the phases due to gravitational interaction and existence of the accelerated region $B$, one gets:
\begin{equation}
    |\Psi_{in}\rangle \rightarrow |\Psi_{out}\rangle= \alpha|\Psi_{0}\rangle + \beta e^{i\phi_{ca}}e^{i\phi_{ac}} |\Psi_{1}\rangle
\end{equation}
where $|\Psi_0\rangle$ denotes the state associated with the outgoing path from the interferometer, $|\Psi_1\rangle$ with the looped path and $|\Psi_{out}\rangle$ denotes the total output state of the beam splitter (i.e. superposition of two output ports).
Replicating this procedure up to the n-loop process we get a partial sum of the geometric series:
\begin{eqnarray}
     |\Psi_{out}\rangle\nonumber &=& \alpha (\frac{1-\beta^{2n}}{1-\beta^2U^2})^{\frac{1}{2}}|\Psi_{0}\rangle + \beta^n U^n   |\Psi_{1}\rangle \\
\end{eqnarray}
where $U=e^{i\phi_{ca}}e^{i\phi_{ac}}$ and $\beta^2=1-\alpha^2$ for normalization in our model. Thus, within the n-looped process there is still non-zero probability of finding a particle within the loop which is not the case in the infinite regime, i.e. this looped interferometer is fully reflective.

For the convergence of the series for $n\rightarrow +\infty$, it is required to put  unitarity of the loop. For the $\infty$-loop process, the series converges to:
\begin{equation}
  |\Psi_{out}\rangle \rightarrow \frac{\alpha}{(1-\beta^2 U^2)^{\frac{1}{2}}}|\Psi_{0}\rangle
\end{equation}
and as expected, this system is reflective.
Noticeably, $\det(1- \beta^2 U^2)\neq 0 $ which results from $|\beta|<1$ (iff $|\beta|=1$ then there is effectively no history branch with the particle outgoing the loop) and unitarity of U (implying  $|U|=1$). This condition leads to convergence of the series. Conversely, if the loop was not unitary and the equivalence principle did not hold, then the series would be divergent and would lead to inconsistency of information processing.
For general variants of the unitary operations replacing the beam splitter and the loop, one can apply the methods engaged in \cite{Czachor}.
This analysis naturally resembles  Feynman diagram methods, however, in our case we consider a space-time loop-diagram.

For the mass particle traversing this looped interferometer we can modify the model of the Colella-Overhauser-Werner (COW) experiment \cite{COW} superposing a state of single neutron in two different positions (heights) of the Earth's gravitation field.
Let us consider a particle of mass m at state $|\Psi\rangle=\frac{1}{\sqrt{2}}(|0\rangle + |1\rangle)$ with the branch $|1\rangle$ getting into the loop of our interferometer. For calculation of the action for the mass particle traversing the arm of an interferometer in the gravitational field $g$, the Lagrangian is straightforward:
\begin{equation}
L_{g}=\frac{1}{2}m(\dot{\lambda})^2-mg\lambda
\end{equation}
and for the arm accelerated with the acceleration rate $a$:
\begin{equation}
L_{a}=\frac{1}{2}m(\dot{\lambda} + at)^2
\end{equation}
Assuming that acceleration of the region B happens with the same value $g$, one gets the gauge parameter $G(x,t)=-mg\lambda t-\frac{1}{6}mg^2t^3$ and the relation between Lagrangians:
\begin{equation}
L_{g}=L_{a}+\partial_{t}G(\lambda,t)
\end{equation}
leading to the output state from the loop interferometer for $n\rightarrow \infty$:
\begin{equation}
|\Psi_{out}\rangle \rightarrow \frac{1}{(2-U^2)^{\frac{1}{2}}}|0\rangle
\end{equation}
where the total $U=e^{i\phi_{a}}e^{i\phi_{g}}$, $\phi_{g}=\frac{1}{\hslash} \int_{t_1}^{t_2} L_{g}d t$ and $\phi_{a}=\frac{1}{\hslash}\int_{t_3}^{t_4} L_{a}d t$.
It is also worth mentioning that the gauge transformation introduces naturally a phase shift $\Delta\phi$ \cite{Marletto} between the two regions A and B. If the proper time of traversing the region A and B is the same, then the difference can be formulated in terms of the gauge transformation, i.e. $\Delta\phi=G(\lambda,t)/\hslash$.

If the equivalence principle did not hold for the quantum systems, then firstly the looped evolution would be not unitary and secondly, we could violate the conservation of energy for the looped  system increasing energy ad infinitum.

It is also interesting to analyze the role of the equivalence principle in quantum algorithms where parts of the algorithms are processed in different gravitational setups. As an example, we consider the Deutsch-Jozsa algorithm represented in terms of the binary tree which has a direct representation by means of Feynman paths. In the following scheme the equivalance principle can play a destructive role in each iteration of the binary tree. If the overall process is not corrected at each step of the algorithm, due to the phase shift acquired by the accelerated branches of the tree, then the final outcome of the algorithm is incorrect. The Deutsch-Jozsa algorithm is initiated with an input register of $n=log_2 N$ qubits (where N represents a binary tree level in our scheme - Fig. 2) in a state $|\Phi_{in}\rangle=|00\ldots 0\rangle$. Then each qubit is processed with a Hadamard operator leading to a superposition of all states in the computational basis of $\mathcal{H}=(\mathbb{C}^2)^{\otimes n}$. In the binary tree approach the whole input register state can be generated by $N$-level tree splitting the state (e.g. by the bean-splitter in the interferomter) at each level of the tree.
The proposal of implementation of such an evolution with the ring cavity is discussed in \cite{DJ, Walk} but it can be also implemented by nested interferometers as aforementioned for the COW interferometer above.
\begin{figure}[h]
\centerline{\includegraphics[width=9.5cm]{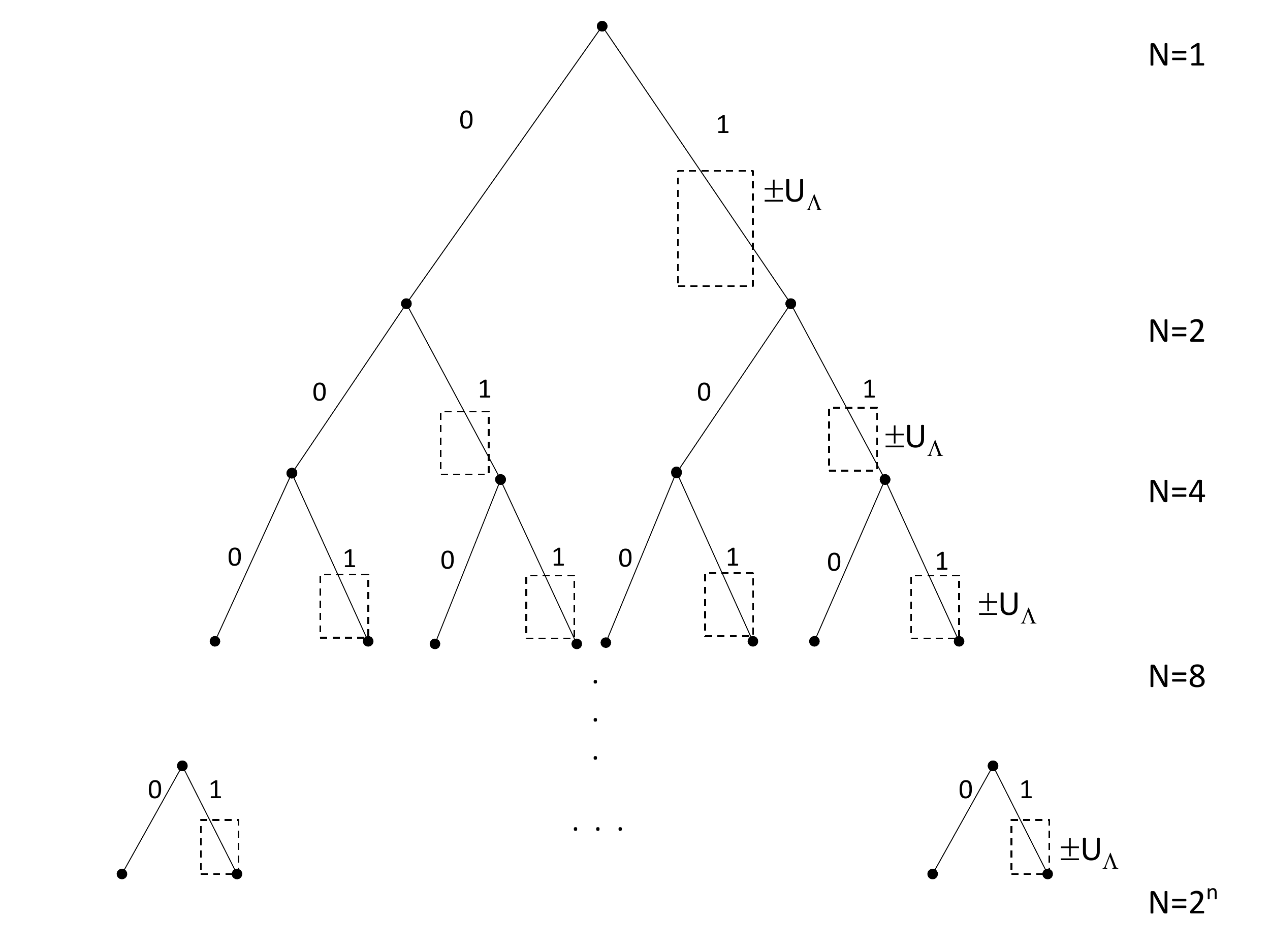}}
 \caption[BinaryTree]{Any binary balanced or constant function can be encoded as a binary tree with the superposition state $\frac{1}{\sqrt{2^n}}\sum_{x=x_1,\ldots, x_n=0,1}(-1)^{f(x)}|x\rangle$ as a sum over all paths up to the N-level nodes. The phase shifts are encoded in the 1-branches. $U_{\Lambda}$ introduces additional gauge phase shift in the process disturbing flow of the algorithm.}
 \label{BinaryTree}
\end{figure}
The process of creating n-qubit paths works as follows:
\begin{eqnarray}
|0\rangle &\rightarrow& \frac{1}{\sqrt{2}} (|0\rangle + |1\rangle) \rightarrow\ldots \rightarrow \\\nonumber
&\rightarrow&\frac{1}{\sqrt{2^n}} \sum_{x=x_1,\ldots, x_n=0,1}|x\rangle
\end{eqnarray}
thus, we consider a N-level tree for creation of the $n=log_2 N$ qubit state (Fig. 2). There are two other fundamental steps in the Deutsch-Jozsa algorithm. The oracle $U_f$ implements the function as constant or balanced which is queried by the whole algorithm:
\begin{equation}
    U_f \frac{1}{\sqrt{2^n}} \sum_{x=x_1,\ldots, x_n=0,1}|x\rangle = \frac{1}{\sqrt{2^n}}\sum_{x=x_1,\ldots, x_n=0,1}(-1)^{f(x)}|x\rangle
\end{equation}
The second step includes again the Hadamard operation and measurement in the computational basis. If the state $|00\ldots0\rangle$ is achieved (with probability $P(|00\ldots0\rangle)=|\frac{1}{\sqrt{2^n}} \sum_{x=x_1,\ldots, x_n=0,1}(-1)^{f(x)}|^2$), then the function is constant. For the balanced function we get destructive interference and the probability of getting $|00\ldots0\rangle$ tends to zero.

In our scheme we engage a binary tree that can encode effectively any binary balanced function $f(\cdot)$ by introduction of the phased paths for the $|1\rangle$-branches (the chosen branches should face a $\pi$-phase shift for the oracle implementation). We can disturb the oracle implementation by introduction of a gravitational phase shift if branches $0$ and $1$ are localized in two different gravitational setups as discussed in this paper:
\begin{eqnarray}
&U_{\Lambda}& \frac{1}{\sqrt{2^n}} \sum_{x=x_1,\ldots, x_n=0,1}(-1)^{f(x)}|x\rangle = \\ \nonumber
&=&\pm \frac{1}{\sqrt{2^n}} (|0\rangle\pm e^{i\phi_g}|1\rangle)(|0\rangle\pm e^{i\phi_g}|1\rangle)\ldots(|0\rangle\pm e^{i\phi_g}|1\rangle)
\end{eqnarray}
where the unitary operator $U_\Lambda$ introduces a $e^{i\phi_g}$ shift (that can be the gauge phase shift due to the equivalence principle) to $|1\rangle$ branches and can effectively change the final answer about the nature of the function $f(\cdot)$. If one considers the scenario with regions A in the gravitational field and B with the corresponding acceleration, then only the difference in phases between the $|0\rangle$ and $|1\rangle$ branches influences the final result of the algorithm - the global phase can be omitted.

In summary, we showed that the quantum version of the equivalence principle guarantees unitarity of quantum information processing in the weak gravitational field and is necessary for keeping consistency of quantum operations, especially if the processing algorithm is distributed spatially across regions with different gravitational  characteristics.
It is an open question how the presented analysis should be modified to be correct also for strong gravitational fields.

It would be also interesting to perform an experiment with light signals exchanged in a looped system between the system A being close to the Earth surface and the system B being on a stationary orbit around Earth to check the presented model.


\end{document}